\documentclass{elsarticle}
\usepackage[utf8x]{inputenc}

\providecommand{\tightlist}{%
  \setlength{\itemsep}{0pt}\setlength{\parskip}{0pt}}

\usepackage{placeins}
\usepackage{mdframed}
\usepackage[hyphens]{url}
\usepackage{lineno,hyperref}
\modulolinenumbers[5]

\journal{Futures, Copyright waived (CC0)\hspace{.7em}\raisebox{-.5ex}{\includegraphics[height=2.8ex]{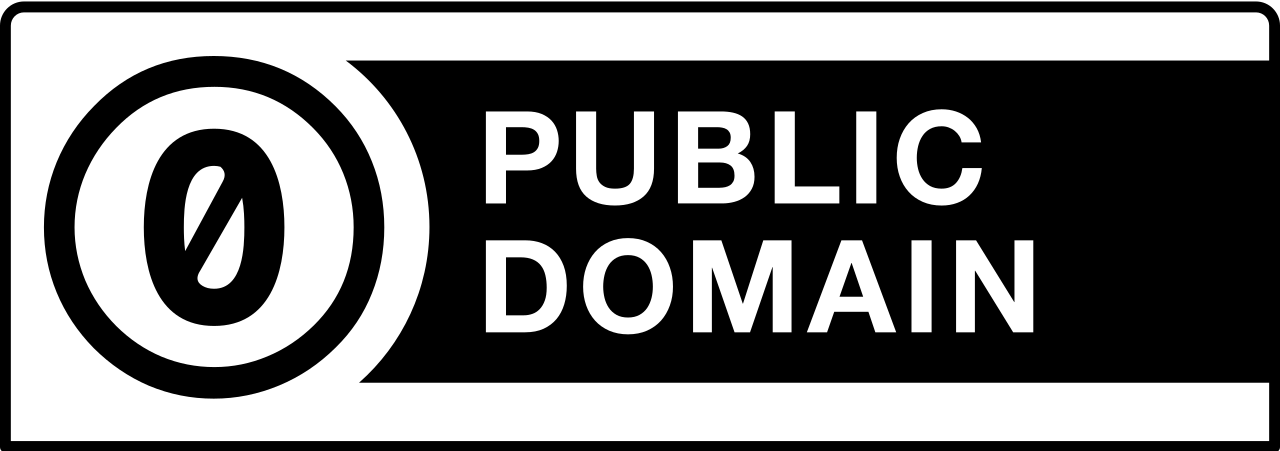}}}

\biboptions{authoryear}

\begin{document}

\begin{frontmatter}

\title{Patterns, anticipation and participatory futures}

\author{Raymond PUZIO\corref{mycorrespondingauthor}}
\address{Hyperreal Enterprises, Ltd. 114A New Street, Musselburgh, Scotland, EH216LQ, UK}
\ead{rsp@hyperreal.enterprises}
\cortext[mycorrespondingauthor]{Corresponding author}
\author{Paola RICAURTE​}
\address{Tecnol\'ogico de Monterrey, Calle del Puente 222 Col. Ejidos de Huipulco, Tlalpan,
14380 Mexico City, Mexico}
\address{Berkman Klein Center for Internet \& Society, 23 Everett Street, 2nd Floor, Cambridge,
MA 02138}
\ead{pricaurt@tec.mx}
\author{Charles Jeffrey DANOFF​}
\address{Mr. Danoff’s Teaching Laboratory, PO Box 802738, Chicago, IL 60680}
\ead{contact@mr.danoff.org}
\author{Charlotte PIERCE​}
\address{Pierce Press, PO Box 206, Arlington MA 02476}
\ead{charlotte@piercepress.com}
\author{Analua DUTKA-CHIRICHETTI​}
\address{DeepResiliency.io, Hauptstrasse 28, 67475 Weidenthal, Germany}
\ead{analua.chirichetti@gmail.com}
\author{Vitor BRUNO​}
\address{Milestone English Course, Rua Trieste 170, ap2, Palhoca, SC, Brazil, 88132-227}
\ead{chief@milestoneenglishcourse.com}
\author{Hermano CINTRA​}
\address{Instituto Kapok de Inovação Corporativa, Rua Estados Unidos, 411, São Paulo, SP,
Brazil, 01457-00}
\ead{hermano.cintra@institutokapok.org}
\author{Joseph CORNELI​}
\address{Hyperreal Enterprises, Ltd. 114A New Street, Musselburgh, Scotland}
\address{Institute for Ethical AI, Oxford Brookes University, Headington Campus, Oxford, OX3 0BP UK}
\ead{jcorneli@brookes.ac.uk}

\begin{abstract}
Patterns embody repeating phenomena, and, as such, they are partly but not fully
detachable from their context. ‘Design patterns’ and ‘pattern languages’ are established
methods for working with patterns. They have been applied in architecture, software
engineering, and other design fields, but have so far seen little application in the field of
future studies. We reimagine futures discourse and anticipatory practices using pattern
methods. We focus specifically on processes for coordinating distributed projects,
integrating multiple voices, and on play that builds capability to face what’s yet to come.
One of the advantages of the method as a whole is that it deals with local knowledge
and does not subsume everything within one overall ‘global’ strategy, while
nevertheless offering a way to communicate between contexts and disciplines.
\end{abstract}


\end{frontmatter}

%\linenumbers

\section{Introduction}

In the midst of the Covid-19 pandemic, it is natural to ask: could we
have been able to collectively anticipate the spread of zoonotic
diseases, understand possible reactions, and formulate more adaptive
responses? Many individuals anticipated that something like this was
coming
\citep{yong2018a,karesh2012a,schwartz1996a,sardar2010a,sardar2016a}. Our
collective struggles have shown that we need to improve our
anticipatory capabilities, that is, \emph{the capacity to act in
  response to or in preparation for a potential future reality}
\cite{unknown2018a}. As the virus spread from one corner
of the world to another, it was no longer simply a question of putting
a stop to the disease, and not simply a biomedical issue. Pandemics
belong to a class of complex problems with social, political, and
economical layers. While we are not going to solve the pandemic in
this paper, the complex challenge motivates the issue we want to
explore: how can communities learn to better anticipate together.

Of course, people have been finding new ways to think about the future
together for a long time. For example, \citet{ariyaratne1977a} tells the
story of a rural group who were finally able to complete an important
construction project. After 15 years of deadlock spent waiting for
outside investment, they were called to a community meeting where they
figured out that they could do the job with their own labour. As
dialogue and inquiry gave participants new ways to articulate and
develop their thinking together, the nature of the problem they faced
became easier to understand and resolve.

The methodology we will develop here centres on \emph{design patterns}:
succinctly, solutions to problems arising in the context of a repeatable
activity. We propose to use design patterns to structure
anticipatory peer learning as a way to relate to possible future
scenarios. The link between design and the future has been aptly
summarised: ``Design is future creating and not future guessing''
\citep[p.~47]{banathy1991a}. However, the relationship between the
well-established design pattern methodology and futures studies has seen
relatively little previous development.

After defining and giving examples of how design patterns have been used
in other fields, we then examine how they can help create peer learning
experiences to explore and prepare for the future. We outline a process
of Open Future Design that we elaborate with three patterns that can
be applied to collaboratively anticipate the unknown:

\begin{itemize}
\tightlist
\item
  Create a roadmap as a tool to guide collaborative creative work.
\item
  Use scenario planning as a quasi-democratic approach to organise
  thinking about the future.
\item
  Play games in a collaborative manner to explore possible futures.
\end{itemize}

Our results present a synthetic, literature-based, treatment of these
topics by way of design patterns. Whether the future holds a
catastrophe, or something much better, patterns can help us evolve in
our collective awareness today.

\subsection{Background}

Life itself is intrinsically anticipatory \citep{poli2012a,poli2020a} and
prefigurative schemas are relevant to the way we learn \citep{spiro1996a}. However, as a society, we are not guaranteed to be able to
produce a viable outcome out of disparate individuals' capacities to
think about the future. How the challenges we face are understood
depends very much on who is brought to the table, and how their
discussions are organised. We have chosen to focus on pattern methods
here because they seem to offer the prospect of being useful for
developing viable, thriving, collective outcomes.

As Covid-19 rages out of control in some places, it is easy to forget
that humans have largely eradicated or successfully controlled other
diseases (e.g., polio, ebola, leprosy). While design patterns do not
directly prevent the spread of disease, they can help prevent uncareful
thinking --- the consequences of which have included failing to
anticipate a problem (as happened in many places with Covid-19 in
December 2019 to February 2020) --- or becoming too complacent based on
previous successes (many people regularly choose not to get a flu shot).
Patterns have been used to support and structure discourse around
complex technical systems, and we feel they have much to offer in a
futures studies context.

Reflecting on the history of design-pattern discourse, \citet{shalloway2005a} remark that the related notion of a `cultural pattern' has
received attention in anthropology \citep{benedict1934a,benedict1946a}. Likewise,
Joseph Campbell proposed the monomyth of the hero as an `archetypal
pattern' that is embedded in myths and stories across diverse cultures
and historical periods \citep{campbell1949a} and Inayatullah noted that
historians have understood historical trends in terms of patterns
\citep{inayatullah1998a}. `Design patterns' in a general sense have long
supported both small-scale and mass-manufacture of material objects and
structures, ranging from clothing and ceramic ware to city-scale
planning and, arguably, beyond.

Notwithstanding these precedents, the concept of patterns with which we
are concerned here originated with the architect Christopher Alexander
and his collaborators in the 1960s and 1970s \citep{alexander1977a}.
One feature of their proposed pattern-based approach was to enable all
stakeholders to participate in the process of architectural design.
Additionally, they promoted a natural quality found in traditional
architecture, versus the artificial one due to central planning
\citep{alexander1965a}.

Alexander's notion of an architectural pattern language was later
adopted and adapted within the field of computer programming
\citep{beck1987a,gamma1994a}. This trend received a large boost with
the invention of wiki software, which was first used to collect and
curate collections of such patterns \citep{cunningham2013a}.
Subsequently, `design pattern' methods roughly in the Alexander style
have been applied to other domains, including social activism
\citep{schuler2009a}, the transition movement
\citep{hopkins2010a,transitionnetworkorg2010a}, organizational
development \citet{manns2015a}, disaster prevention
\citet{furukawazono2013a}, and mental health
\citet{pierri2016}. Particularly relevant to our interests,
\citet{schuler2009a} speaks about patterns as conceptual tools
for the future, and
developed a pattern called
``\href{http://www.publicsphereproject.org/node/543}{\emph{Future
    Design}}.'' Iba develops a future-oriented view on pattern
language discourse, with a recent presentation on ``Creating Pattern
Languages for Creating a Future where We Can Live Well'' \citep{iba2019a}.

Since the definition of a pattern has been criticized for being
inexplicit and inadequately explained \citep{dawes2017a} and
treatments by different sources vary, we would like to begin by
explaining our understanding of the term \emph{design pattern}. We
recognise that some readers may bring their own existing understandings
of this term, and that some will have no previous familiarity with it at
all.

As our starting point, we take Alexander's phrase: ``Each pattern is a
three-part rule, which expresses a relation between a certain context, a
problem, and a solution'' \citep[p.~247]{alexander1979a}. However, 
\citet{gabriel2002a} emphasises that Alexander's full description of patterns goes
well beyond the one sentence quoted above. Leitner supplied the
following further summary text: ``Patterns are shared as complete
methodic descriptions intended \emph{for practical use by experts and
non-experts}'' \citep{leitner2015a} {[}emphasis added{]}.

As we look into the matter two central elements emerge. Like an ellipse,
the concept of the design pattern has two foci: context and community.
The context is the type of activity which is being considered, such as
designing a building or writing software. Communities include
stakeholders --- experts and non-experts alike --- who are involved with
or affected by a particular project. Also central to Alexander's
pattern-based methodology is the notion of a \emph{pattern language}, in
which each pattern description itself is contextualised by related
pattern descriptions that deal with related problems. By working
together as a whole, a pattern language produces coherent entities
\citep{alexander1999a}. Within a specific design challenge, cascading
patterns overlap to build out a design solution: for example, to fully
develop a building, one would require a pattern for a suitable type of
roof.

A pattern identifies how a community may proceed to obtain an outcome by
making certain choices and applying specific methods to suitable
materials. At the heart of the design pattern method is the feature of
practicality. Each pattern description, when followed, should help the
people using it --- the community --- overcome some real or potential
conflict \citep[pp.~9-10]{alexander1970a}.

We will argue that patterns can be used to promote the organic, vital
quality of resilience, that is, ``not to be well-adapted, but to adapt
well'' (\citet{downing2007a}; quoted in \citet{tschakert2010a}). We find
the `non-expert' portion of Leitner's definition particularly important
and hope to explain our ideas in an accessible way.

The most tangible metaphor is to treat each pattern as a journey, ``a
path as a solution to reach a goal'' \citep{kohls2010a,kohls2011a}. In this
treatment, patterns are understood to have an initial condition and an
end condition, defined within some context, and the corresponding
problem is understood to be that of finding a good way to move from the
initial state to the goal.

The economist Elinor Ostrom related Alexander's pattern
language to Arthur Koestler's notion of `holons' --- stable components ``in an
organismic or social hierarchy, which displays rule-governed behavior''
(\citet[p.~11]{ostrom2009a}; \citet{koestler1973a}). This citation makes explicit the analogy between
patterns and Ostrom-style institutions. Ostrom's framework understands
an institution as a stable set of action-situations and proposes that
each such situation be understood as a whole, while simultaneously being
part of a larger social arrangement and a result of particular
circumstances. In short, pattern language and institutional design are
seen to be closely compatible.

In their applications within the computing discipline, patterns are
typically written down with a template for easy communication and
discussion. For example, alongside ``Context'', ``Problem'', and
``Solution'', a typical template may include things like ``Rationale''
(reasons for preferring this specific solution over other possible
solutions; Meszaros \& Doble, 1997) and ``Examples''. Typically a
picture is included to serve as a mnemonic. Corneli et al.~(2015) added
a ``Next Step'' facet to the template they used, with the implication
that each pattern is imperfectly realised and that further steps should
be taken to manifest it within a collaboration.

Patterns have also been discussed in explicitly computational terms,
though that direction of work remains mostly at the level of a proposal
\citep{alexander1999a,moran1971a}, which has seen limited
discipline-specific uptake within architectural design
\citep{jacobus2009a,oxman1994}. As a whole, the pattern discipline continues to evolve as
an active area of research, particularly in computing disciplines.
Several parallel yearly conferences are devoted to the exploration of
themes touched on here, with some of the most developed contributions
appearing in the Transactions on Pattern Languages of Programming book
series. The specific contribution we make in this paper is to unite the
pattern language and futures discourses as a coherent whole.

\subsection{Pattern(s) as anticipatory method}

In an everyday sense, `patterns' are elements of an environment that
recur in time or space. Because we expect them to recur, the patterns we
recognise bear on our thinking about the future. The methodology we
apply below is the reanalysis of existing future-oriented discourse
within the specific constraints of the design pattern approach. The goal
is to reimagine and reassemble it ``together as a different version to
bring out something latent or implicit in the original''
\citep[p.~85]{handelman1998a}.

As we reviewed above, we refer here to a specific methodological system
centred on pattern descriptions that are intended for practical use
\citep{leitner2015a}. Patterns of this form are accompanied by and embody
principles that can lead from problem identification to problem solution
within a given domain.

In concrete terms, we imagine how existing future-oriented discourses
can be discussed and thought about using the pattern format. If the goal
of future planning is to move from being reactive to being proactive,
patterns methods have much to offer: they can help imagine and
anticipate possible futures, and can help communities build resilience.

Through the comparative analysis of different future-oriented discourses
vis-à-vis design patterns, we reimagine patterns as being something more
than design tools: they become the building blocks for possible futures.
Used in this way they present a way to ``deconstruct {[}...{]}
conventional thinking to produce a shared view of possible future
outcomes that can break existing paradigms of thinking and operating''
\citep{euforesight2020a}.

From the many existing approaches to scaffolding
anticipation---including speculative fiction
\citep{gill2013a,liveley2017a,wagner-lawlor2017a}, design fiction
\citep{danoff2019,kerspern2019a,di2019a} prospective studies
\citep{jouvenel2004a} and
predictive statistical models \citep{arslan2006a}---we get limited
insights about how to engage together. It is understood that
anticipatory behavior can be provoked and improved \citep{bishop2018a}; and
as individuals, we can engage in anticipatory learning, ``the general
mechanism of learning to generate predictions or learning a predictive
or forward model of an encountered environment or problem''
\citep{butz2012a}. We see design patterns as filling a gap that addresses
participatory anticipatory learning for complex problems requiring a
group response.

\section{Open Future Design: Patterns for anticipating the future}

In this section we use design patterns to analyze three approaches to
thinking about the future, namely: \emph{roadmaps}, \emph{scenario
planning}, and \emph{play}. We posit that the process of anticipation
requires a set of choices that will be associated not only with our
vision of the future --- values, situations, materialities --- but also
with the process in which this anticipation happens.

To begin our exploration of how design patterns relate to futures
studies, we refer to \citet[Appendix, pp.~241-248]{schwartz1996a}, \emph{viz.},
his ``Steps to Developing Scenarios''. This process follows an outline
with a striking similarity to a design pattern template. Both Alexander
and Schwartz advocate the identification of driving forces in a context.
However, unlike Alexander, Schwartz does not intend to resolve conflicts
between the forces within a harmonising design. On the contrary, the aim
in scenario development method is to understand how these forces might
evolve and lead to diverse scenarios. As scenarios develop, they can
serve as the ground for developing new design work in Alexander's sense.
With these aims in mind, the pattern Open Future Design serves as an
entrypoint (Figure 1). As a mnemonic image, we chose to illustrate this
pattern with a depiction of the water cycle, to show how Open Future
Design, and subsidiary patterns that we describe below, represent an
anticipatory cycle: as unknown and unexpected features materialise, our
expectations develop and change.

\begin{figure}[h]
  \begin{center}
    \begin{mdframed}
        \begin{center}
          \includegraphics[width=.9\columnwidth]{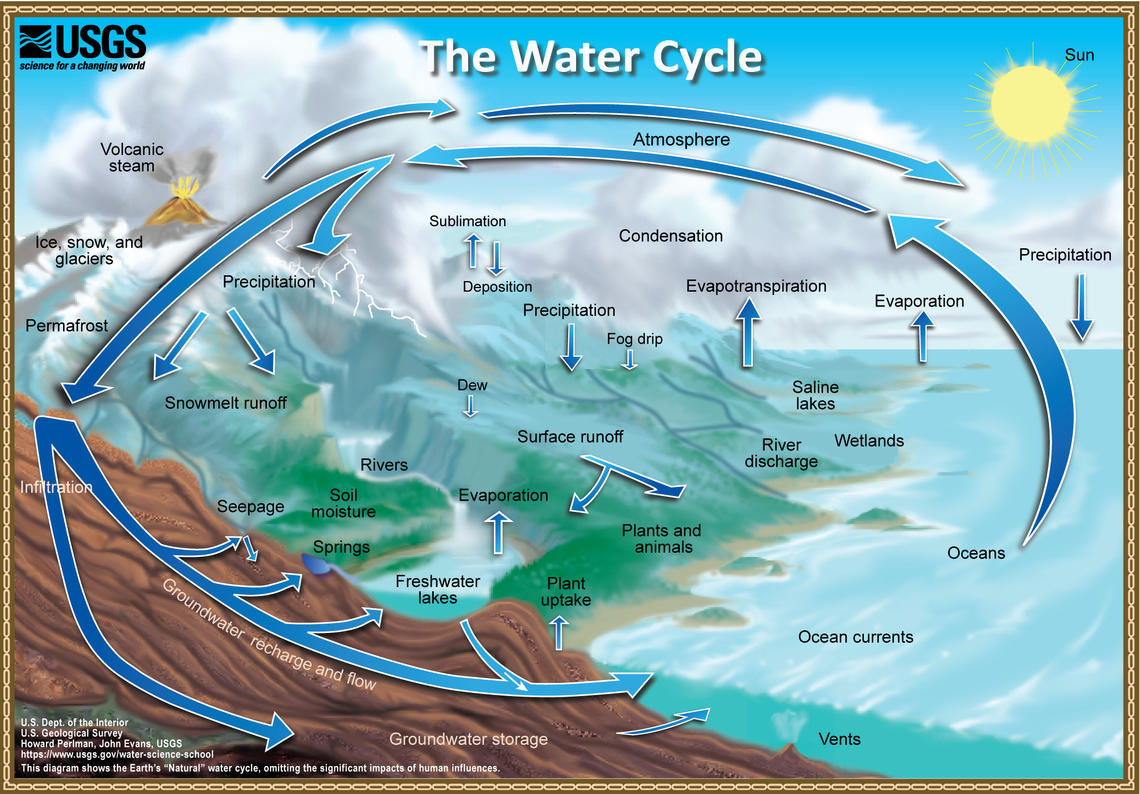}
         \end{center}
\begin{minipage}[t]{0.97\textwidth}\raggedright
\subsection*{{\upshape\texttt{PATTERN}}: Open Future Design}
\smallskip

\textbf{Context} People need to coordinate, plan, and maintain social cohesion.\newline
\textbf{If} a culture can develop based on shared learning BUT there is no reliable oracle that can tell us what to expect;\newline
\textbf{Then} use design pattern methods to articulate multiple futures. This work can be guided by further patterns, e.g., to develop
\medskip

- \emph{a language for projects} → Roadmap

- \emph{a language of future scenarios} → Participatory Scenario Planning

- \emph{a language of roles} → Play to Anticipate the Future
\end{minipage}
\end{mdframed}
\end{center}
\caption{Open Future Design Pattern. Inner image credit: Howard Perlman,
USGS. Public domain, Source:
\href{https://www.usgs.gov/media/images/water-cycle-natural-water-cycle}{\emph{https://www.usgs.gov/media/images/water-cycle-natural-water-cycle}}}
\end{figure}

\subsection{The future via a roadmap}

The success of collaborative learning and work is related to the
conditions, processes, and tools that make such collaborations possible.
One indicative example of `success' is found in the Wikimedia Foundation
(WMF). The WMF has made use of various planning exercises, ranging from
the collaboratively produced ``Wikimedia Strategic Plan'' \citep{walsh2011a}
to the more centrally produced ``Knowledge Gaps'' \citep{zia2019a}
whitepaper. These activities should be understood in the context of an
ongoing decline in participation in Wikipedia projects \citep{simonite2013a},
and the capture of online attention in various privatized media \citep{fuster2011a}.

WMF's strategy and plans are not dissociated from other practices in
their projects. Structured guidance is made available to contributors to
Wikimedia projects at all stages. Such guidance ranges from policies to
tutorials, and helps contributors produce and maintain content to the
benefits of the entire user community. Specific guidance on how to
participate in data analysis and other planning activities that can help
steer the project is available.

The culture around maintaining, updating and using such anticipatory
artifacts comprises a system for governance. For example, online
governance may evolve along the bottom-up lines proposed by
\citet{schneider2020modular}. In the WMF example, the context, problems, and solutions
are relatively explicit --- e.g., gaps in knowledge should be filled to
make Wikipedia more inclusive --- and the community is included in both
framing the problem and shaping the solution. On the other hand,
justifiable doubts are likely to arise among constituencies who are
brought into a ``culture of consultation'' \citep{featherstone2017a} that is
concerned, not with authentic community power, but with establishing and
rubber-stamping the semblance of participation.

One way to enhance community power is to make plans based on pattern
methods. In what might be a best case scenario, the ``Roadmap'' pattern
from \citet{corneli2015a} becomes the entry point to a ``living
language'' made up of other evolving patterns \citet{alexander1977a},
each pointing to clearly designated next steps. While any notion of a
plan presumes an anticipation of outcomes, the Roadmap may not be linear
and directed towards one goal. It may instead be an organic tool for
coordination across variables such as time, groups of people, and
software. In the Scrum agile project management methodology, the
``Product Roadmap'' is complemented by a ``Product Backlog'' that offers
a best-guess linear path through the various branching paths contained
in the roadmap, while the roadmap itself keeps track of all anticipated
alternatives as well as dependencies between tasks
\citep[p.~220, pp.~222-223]{sutherland2019a}.

We can consider the pheromone trails of ants as an example of an
adaptive Roadmap in nature. By analogy, human participants in a
collaborative design process should, metaphorically, have their antennae
out to detect and respond to developing situations, and should develop
suitably immediate methods for sharing meaning with each other. Notice
that by anticipating future conflicts, we can take steps in advance to
prevent and mitigate the problems that might ensue. This connects the
Roadmap pattern with the broader themes of improving our ability to
anticipate the future and resiliency. Figure 2 re-summarizes the Roadmap
pattern, and highlights its nested context, in which the immediate
community of users must adapt to a broader context in motion.

\begin{figure}[h]
  \begin{center}
    \begin{mdframed}
      \begin{center}
\includegraphics[width=.9\columnwidth]{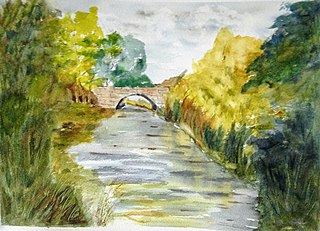}
        \end{center}
\begin{minipage}[t]{0.97\columnwidth}\raggedright
  \subsection*{{\upshape\texttt{PATTERN}}: Roadmap}
  \smallskip
  
\textbf{Context} a group needs to coordinate its activities over a period of time.\newline
\textbf{If} the landscape is complex and not completely knowable BUT adjustment to action based on feedback is possible;\newline
\textbf{Then} use an explicit mechanism to share information about goals, obstacles, methods, and resources.\newline\smallskip

\textbf{Example} Everyday roadmap languages include both iconic map and road sign symbols; when people are confused or lost they may ask for help or try to find their own way back to the road using other informal languages.

\end{minipage}
\end{mdframed}
\end{center}
\caption{Roadmap pattern. Image: Maritess Sulcer, CC0, via Wikimedia
Commons
\href{https://commons.wikimedia.org/wiki/File:Trees_and_bridge_and_stream_and_clouds_--_25_of_33.jpg}{\emph{https://commons.wikimedia.org/wiki/File:Trees\_and\_bridge\_and\_stream\_and\_clouds\_-\/-\_25\_of\_33.jpg}}}
\end{figure}

\FloatBarrier

\subsection{The future via peer learning and scenario planning}

While scenario planning can be carried out by groups of experts, there
is the danger that the life experiences and opinions of other people who
are affected by their planning will not be considered. This could lead
to a narrow understanding of the problem and a limited vision of what
the future can look like. In \emph{participatory }scenario planning, the
activities are carried out by a community of non-experts with the
experts or designers limiting their role to facilitation and moderation
\citep{johnson2012}. For instance:

\begin{itemize}
\tightlist
\item
  The Consultative Group for International Agricultural Research project
  in Mali held a workshop at which 38 people including farmers,
  journalists, local leaders, forest officers, extension workers, policy
  makers, and representatives of NGOs and District Councils collaborated
  to produce scenarios for the future of agriculture \citep{johnson2012,totin2018}.
\item
  Agricultural development workshops in Nigeria included a
  teacher/researcher, a journalist, an engineer, an animal nutritionist,
  a camera operator and head of a women's group \citep{olabisi2016a}.
\item
  In the Minnesota 2050 project, participants were selected from a
  variety of professions and leadership roles to produce scenarios for
  energy and land use, and combined modelling with scenario planning
  \citep{olabisi2010}.
\end{itemize}

In all of these examples, scenario planning was taken out of the hands
of the experts, but scientists and other consultants still played a
specialized role. A varied community that designs and produces something
by and for themselves would be more fully analogous to Alexander's
vision of community-designed architecture \citep{alexander1977a}. If
community members could be involved as scenario planners, or as
architects, why not also involve them as citizen scientists or
modellers? \citet{wildschut2017a} points out that the `local knowledge' of
citizens can make the inquiry process more robust: ``Citizens have
valuable knowledge that is often out of reach for university
scientists.'' Another related people-centered approach
is described by the \citet{designjusticeorg2018a} whose principles are
addressed to ``people who are normally marginalized by design.'' Whether
in a scenario planning, scientific inquiry, or design process,
participants need more than just `access', i.e., not just access to
existing scholarly materials produced elsewhere, but also to a platform
and literacies/capabilities that allow them to contribute and negotiate;
Heidegger used the term \emph{Zuhandenheit}: “readiness-to-hand,
handiness.” Existing collections of design patterns can help address
these requirements. For example, many of the ``Wise Democracy'' patterns
elaborated by the \citet{institute2019a} would carry over well to participatory scenario planning. Specific
patterns include: (\#30) ``Expertise on tap (not on top)'', (\#33)
``Feeling heard'', (\#82) ``Systems thinking'', and (\#94) ``Wise use of
uncertainty''.

The pattern language contains other useful hints for scenario planners.
We can elaborate with a concrete example. \citet{johnson2012} made
this observation based on their experience with participatory scenario
planning: ``Groups with too little diversity converge quickly and
without exchanging contrary ideas, and will necessarily develop fewer
options for action. Yet too much diversity can hobble a group and
overwhelm a process with too many conflicting values and perspectives.''
We do not consider that there is such a thing as ``too much diversity''
in any absolute sense, only limitations to a group's ability to
integrate diverging visions of the future and to learn from conflict.
Indeed, since managing conflict seems to be a key concern here, it would
make sense to express this experience with a design pattern. One benefit
of diversity was noticed by \citet[p.~97]{schwartz1996a}: “being part of
several networks not only opens up diverse arenas, but allows you to
cross-check the insights that emerge among people from vastly different
places.'' Relevant patterns are again found in the Wise Democracy
pattern language: (\#26) ``Diversity'' and (\#88) ``Using Diversity and
Disturbance Creatively''. Scenario planners who are concerned that there
is too little or too much diversity within their dialogue could refer to
these patterns to enrich their thinking on the matter with to-hand
knowledge. An example of where this could have been helpful: \citet{ewing2018a} describes how, when forty-nine individual public schools in
Chicago closed, students were not equal participants in the decision and
were left to deal with the ``intense emotional aftermath'' (p. 159). She
suggested that ``perhaps CPS can learn from the school closures to
create structures that let stakeholders participate meaningfully in
decisions so as to incorporate community voices'' (p. 161).

Patterns could help with further specifics of process. One reference
point in futures studies is the Delphi method \cite{corporation2020a}.
However, in Delphi, summaries are produced by a facilitator synthesising
expert feedback and sharing it back out to participants. Delphi's “goal
is to reduce the range of responses and arrive at something closer to
expert consensus” (\emph{ibid.}). In a world in which we
respect each participant as an expert with relevant experiences, we need
more complex ways to integrate multiple voices. The pattern methodology
itself has the potential to incorporate contributions from diverse
stakeholders, by foregrounding context and community at each stage. It
is worth pointing to extant patterns \emph{for} pattern writing \citep{iba2016a,meszaros1997a} and workshopping
\citep{coplien1997a,gabriel2002a}; and to criteria for evaluating patterns
throughout their lifecycle \citep{petter2010a}. We summarise these
reflections in a ``Participatory Scenario Planning'' design pattern in
Figure 3.

\subsection{The future via play}

Before the future arrives, one way to prepare is by playing: this not
only simulates different scenarios, but also helps us experience what it
is like to live in them. ``Play can help kids learn, plan and even
persevere in the face of adversity'' \citep{willyard2020a}. There can be
other benefits---for adults as well. According to legend, the ancient
Maeonians developed various games (played with dice and balls, as well
as a lottery) that helped them endure hunger \citep[p.~43]{herodotus1975a}.
According to \citet{huizinga1949a}, play is in fact fundamental for
civilization. In a metaphysical view that centers aleatory practice,
``the game and fate fuse into destiny'' \cite[p.~25]{gane1991a}.

\begin{figure}[h]
  \begin{center}
    \begin{mdframed}
      \begin{center}
        \includegraphics[width=.6\columnwidth]{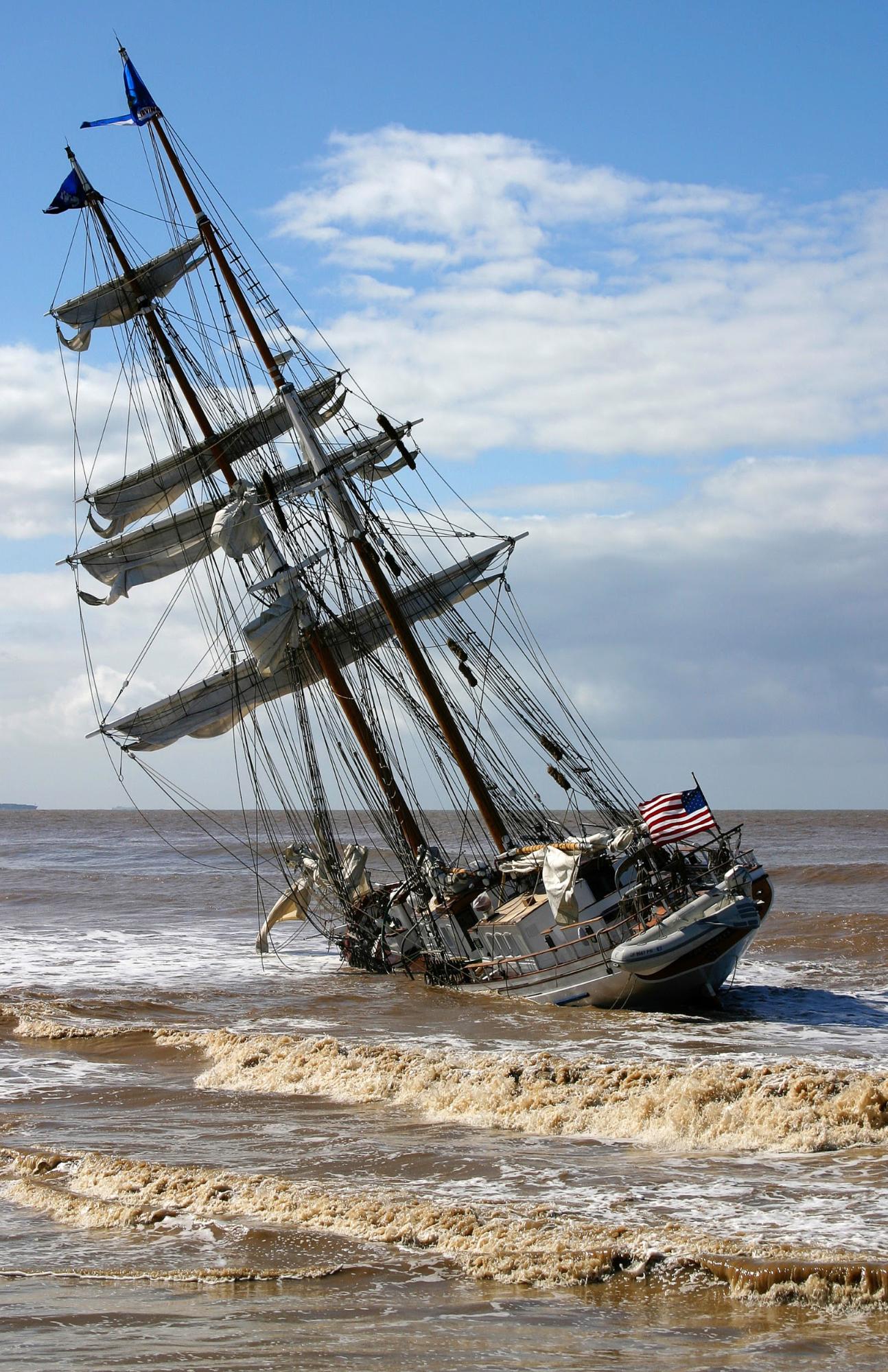}
        \end{center}
\begin{minipage}[t]{0.97\columnwidth}\raggedright

  \subsection*{{\upshape\texttt{PATTERN}}: Participatory Scenario Planning}
\smallskip
  
\textbf{Context} you want to plan for possible future scenarios.\newline
\textbf{If} you have an interested group BUT no “expert” has all the answers;\newline
\textbf{Then} pool the collected expertise of the affected communities.\newline\smallskip

\textbf{Example} the Next Generation Services project (\url{http://nextgenpsf.co.uk}) used participatory scenario planning to imagine different plausible AI futures, which were then used in design sprints to expose professional services firms to potential AI challenges and opportunities.
\end{minipage}
\end{mdframed}
\end{center}
\caption{Participatory Scenario Planning.
\href{https://commons.wikimedia.org/wiki/File:IrvingJohnstonAground.jpg}{\emph{Image}}:
The tall ship\href{https://en.wikipedia.org/wiki/Irving_Johnson_(ship)}{
}\href{https://en.wikipedia.org/wiki/Irving_Johnson_(ship)}{\emph{Irving
Johnson}} lies hard aground, only yards from shore, near the entrance to
Channel Islands
Harbor,\href{https://en.wikipedia.org/wiki/Oxnard,_California}{
}\href{https://en.wikipedia.org/wiki/Oxnard,_California}{\emph{Oxnard,
California}}, March 2005. US Coast Guard, public domain.}
\end{figure}

\FloatBarrier

Games offer `oblique strategies' (à la Eno \& Schmidt --- see \citet{taylor1997a})
for anticipating the future. They transform the future into
something we can play with and thereby expand the horizons of what we
believe is possible in the coming days and years \citep{Sweeney2019}.
This has not gone unnoticed by futurists. For example: in 2015, United
Kingdom government officials developed a futures gaming process called
the ``Scenario and Implication Development''
system to play at the 5th edition of the Global Strategic
Trends review \citep{sweeney2017a}. Causal Layered Analysis (CLA)
\citep{inayatullah1998b} is a well known tool for imagining possible
futures: at the conference “Futures Studies Tackling Wicked Problems”
in Finland, attendees played a CLA game to test out different
scenarios, and the game allowed them to include a diversity of
perspectives \citep{heinonen2017a}.

Examples of recent games designed to explore possible futures include:

\begin{itemize}
\tightlist
\item
  \emph{\textbf{The Oracle for Transfeminist Technologies}} \citep{varon2019a}: a card deck designed to collectively envision and share
  ideas for transfeminist tech in the future. This game is based on the
  assumption that people who are excluded by technology in today's world
  are qualified to design future scenarios \citep{varon2020a}.
\item
  \emph{\textbf{Flaws of the Smart City}} \citep{friction2016a}: a
  card game exploring smart-city futures (that architects like Alexander
  might also enjoy playing). The co-founder of Design Friction explains
  how the game tries to make it easier for those outside the design
  community to easily use future scenarios and design fiction \citep{kerspern2019a}.
\item
  \emph{\textbf{HEY! Imaginable Guidelines}} \citep{sanalarc2018a}: another
  city themed game in which participants use cards to discover how
  cities are designed by selecting topics that are necessary, desirable,
  or irrelevant to the urban-design problem they are trying to solve
  \citep{hattam2019a}. The game provokes players to make changes and continue
  the conversation after play has concluded.
\item
  \emph{\textbf{Equitable Futures}}: a future-oriented game produced by
  the Institute For the Future. It is used to develop understandings of
  challenges to overcome as players try to find new ways to build a more
  equitable future and re-imagine our worlds for 2030 \citep{finlev2019a}.
\end{itemize}

Although it was not designed by futurists, the cooperative board game
\emph{Pandemic} \citep{leacock2008a}, where players work together to find a
cure for a virus spreading across the world, has served a similar
purpose. It has helped people cope with the current Covid-19 pandemic
\citep{genovese2020a,leacock2020a}, including doctors and medical students
\citep{woo2020a,borrelli2020a}. Employees of the Centers for Disease
Control previously commented that the game ``reflected the reality and
values of public health'' \citep{lee2013a}. In this case the upshot is that
players have been experimenting with how to solve a pandemic, thanks to
this game, since 2008.

There are existing design patterns for creating games and using them to
learn. Iba's pattern
“\href{https://books.google.com/books?id=9yJPBwAAQBAJ\&lpg=PA35\&ots=cplh2j7i7i\&dq=takashi\%20iba\%20play\%20game\%20pattern\&pg=PA35\#v=onepage\&q=game\&f=false}{Playful
Learning}” is based on the observation “It is easier to enjoy learning
something new if you take pleasure in the results” \citep[p.~9]{iba2018a}; for
instance, one might opt to learn to program by programming a video game,
or practice a foreign language by playing Two Truths and a Lie. The
genre of “`Applied Games' or `Games with Purpose' ... seek to harness
our natural curiosity and playfulness in the search for new knowledge”
\citep{rawlings2016a}. One sub-genre is described in Schuler’s
``Socially Responsible Video Games Pattern'' where games are used as a
vehicle to introduce real-world issues in an interactive format \citep{schuler2008a}.

As Iba's
``\href{https://books.google.com/books?id=9yJPBwAAQBAJ\&lpg=PA35\&ots=cplh2j7i7i\&dq=takashi\%20iba\%20play\%20game\%20pattern\&pg=PA35\#v=onepage\&q=game\&f=false}{Playful
Learning}'' pattern stated, the games themselves need to be fun and
tangible. As one simple step in this direction, Iba created
physical cards for his patterns, so that playing with them would be more
enjoyable. Collections of game design patterns are also available that
can help designers build in features that make games attractive and
interesting \citep{kreimeier2002a}.

Figure 4 summarises the foregoing reflections as a pattern, Play to
Anticipate the Future. Perhaps future definitions for \emph{play} will
catch up with the reality this description embodies: none of the five
current definitions for this word as a verb in the Webster dictionary
include anything about learning or the future \citep{merriam-webster2020a}.

\section{Discussion}

Our presentation at the Anticipation 2019 conference took the form of a
scripted dialogue that we invited audience members to read aloud. We
hoped that this might spark conversation, by modelling the kind of
participatory dialogue we wanted participants to feel comfortable
having---better than one presenter speaking to the audience the entire
time. As it happened, the first comment we received from one of the
volunteer readers pushed back vigorously against the format: ``Why did
you ask us to read a script, rather than improvise?'' Subsequently, this
led us to explore more authentically collaborative methods for
anticipating the future. Central to this process was the fact that we
are a diverse international group of peers who were prepared to think
together about our local concerns, investments, and activities. This
paper has shown how the pattern method can be used to develop a mindset
for collaborative anticipation that embraces diverse points of view.
Such usage was already hinted at by \citet{moran1971a}, who wrote that ``From
the point of view of methodology, it is not so important how good each
pattern is, but only that each one is transparent and open to criticism
and can be improved over time.''

With further application of the method, futures discourses could become
more `generative', i.e., refashioned as ``a kit of parts \ldots{}
together with rules for combining them'' \citep{alexander1968a}. To meet this
need, the set of patterns would have to be more fully elaborated. In the
domain of the built environment, \citet{alexander1999a} refers to inspiration
coming from ``generative schemes that exist in traditional cultures''
with ``as few as a half a dozen steps, or as many as 20 or 50.''
It is not simply a matter of adding more patterns ---
but one of understanding the unfolding processes that they represent, when taken together.

\begin{figure}[h]
  \begin{center}
    \begin{mdframed}
      \begin{center}
\includegraphics[width=.7\columnwidth]{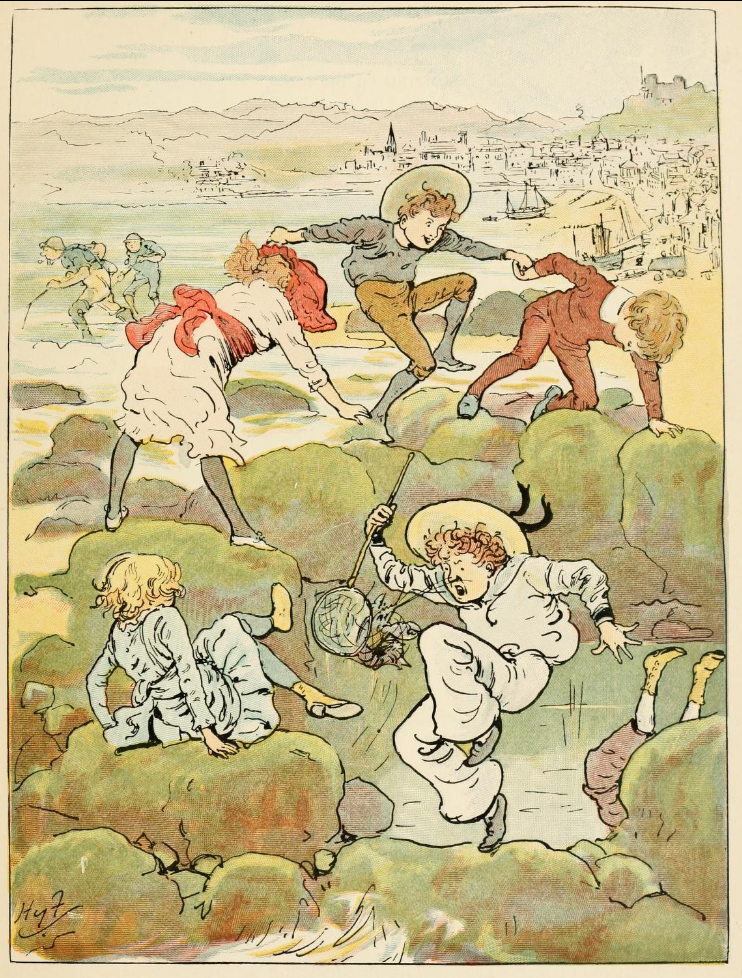}
\end{center}
  \begin{minipage}[t]{0.97\columnwidth}\raggedright
\subsection*{{\upshape\texttt{PATTERN}}: Play to Anticipate the Future}
\smallskip

\textbf{Context} you want to have fun with friends, colleagues or acquaintances.\newline
\textbf{If} you want to explore possible futures BUT time travel does not exist and you don’t know what to expect;\newline
\textbf{Then} play a game that lets you experience a plausible future scenario together.\newline\smallskip

\textbf{Example} a bipartisan group of politicians, former civilian and military officials, and academics gathered to play a scenario planning game to anticipate the possible aftermath of a contested election: at the very least the scenarios they came up with managed to surprise them (Bidgood, 2020).

\end{minipage}
\end{mdframed}
\end{center}
\caption{Fig 4. Play to Anticipate the Future Pattern, illustration from
\href{https://archive.org/details/romps00furn/page/22/mode/2up}{\emph{Romps}}
by Harry Furniss, published by Routledge and Sons in 1886 (public domain
via Archive.org)}
\end{figure}

\FloatBarrier

Pattern language development can give us another way to think
and talk about 
the ``process of language development'' that already goes
on inside institutions \citep[p.~205]{schwartz1996a}. Further challenges
include understanding complex interactions between contexts and
communities that previously had little relationship with each other.
Another concern is how to balance futures discourse with other pressing
concerns: if we wait too long, or structure the discourse poorly, we
will end up being reactive. One specific direction for development
of the ideas would be to work jointly with the Participatory Modeling (PM)
community, addressing concerns like ``What are the major interpretation
and communication issues in PM? Are there best practices to address
these issues?'' \citep{jordan2018a}. To be most applicable in this
setting, the methods would need further elaboration, e.g., to become
more computationally salient to address the specific needs of modelers,
and to incorporate the developing insights from Alexander and others on
what makes this method work well.

The vision of ``Open Future Design'' (§2) is to be open to almost any
possibility when we imagine it. As a touchstone, when they are first
suggested, beneficial future visions ``should appear to be ridiculous''
\citep{dator2019futures}. However, in real life, there are scenarios we prefer to
close off. We can prepare for various eventualities by experiencing
these possibilities in games (§2.2) or other simulated encounters,
rather than in real life.

A pattern-theoretic approach to futures links `futures studies' to
processes of growth and development. Community assembly of ecosystems
and the growth of embryos are relevant natural examples of processes
that have `futures'. These natural processes are ordered, not only in
time but also in space. Reflecting on these examples, we see more
clearly how it is that individual futurists --- without the contexts in
which their ideas could become meaningful --- are like little bits of
organs that are detached from the bodies and materials needed to
function. Similarly, while science fiction can provide a `thinking
machine' \citep{Doherty2020}, it cannot on its own provide
historically robust alternatives to the global crises that we face.
Narrative and design fiction methods \citep{gill2013a,liveley2017a,wagner-lawlor2017a} have been applied in design, and from there,
reapplied in futures studies. When these methods are used to prepare
effectively, we would be inclined to think of them as part of a
``Roadmap'' in the sense we used this term above (§2.1).

A roadmap is a kind of genetic code for a successful project: and here
one should keep in mind that a genotype is not a simple blueprint
\citep{pigliucci2010a}. You cannot create a roadmap in a vacuum and hand it
to someone and expect it to work. Collaboration and communication are
needed to create a useful roadmap and that process itself can become the
largest part of the value of the roadmap. Once the plan is in motion and
things start not to happen as expected, the roadmap can be used as a
reminder of what was planned to help groups respond based on prior
collaborative knowledge and understanding. The understanding it took to
create it is key. It creates “resilience”.

Diversity --- secured via patterns like ``Participatory Scenario
Planning'' (§2.2) --- can boost our collective immunity in a broad sense
of that word \citep{Wambacq2017}. Diversity allows our
understanding of the whole to be enriched. Without it, we risk attacking
potentially vital imagined futures, as when in autoimmune disease, the
immune system attacks healthy body cells \citep{handbookdao}. If
we fail to consider and account for the needs of all stakeholders
affected by decisions, we are not acting in alignment with our basic
need for survival as part of humanity and part of the ecosystem on
earth.

As we consider the needs and interests of broad cohorts of stakeholders
we can position this work as a response to \citet{KOSTAKIS2015126} who
argued for a development model based on ``thinking global and producing
local.'' At the centre of their vision is a global pool of designs,
which are put into production in local Fab Lab facilities. By contrast,
in our work we have centred circumstances of cultural diversity and
human relationships. In our view, the pattern method flips the Kostakis
et al.~formula on its head: patterns are primarily tools for thinking
locally about particular contexts, individual relationships,
conflicts and circumstances. Only secondarily and potentially does this
lead to a shared global resource. More likely, pattern methods would
simply strengthen local forms of resilience and better identify healthy
futures.

Incorporating diverse communities into an ongoing futures discourse is a
tall order and some humility is required. In contrast to design, which
ultimately converges, futures-oriented activities are about producing
diverse open-ended possibilities: this is akin to Deleuze's remark that
``the system must not only be in perpetual heterogeneity, it must also
be a heterogenesis'' \citep{deleuze2010a}. This is why reading a script --- for
example --- is not necessarily the best way to motivate discussion about
the future. As \citet{sardar2010a} pointed out, future predictions do not
actually provide us with foreknowledge. However, anticipation can
provide future-readiness; with this in mind, a script could be reworked
as a game that participants could ``Play to Anticipate the Future''
(§2.3) together.

The project we have embarked on will help to address the societal need
for improved facilities of anticipation. Through the generative nature
of the pattern method, users can become prepared for many outcomes and
possibilities. A great strength of this method is that it can be used
differently in different places, with no need of assembling everything
into one unified `design'. We highlight some specific implications of
the patterns we developed in Figure 5.

The individual points indicated in Figure 5 are at least somewhat
plausible, e.g., Ali S. Khan  from the CDC ``became interested in
Vaughan's game {[}\emph{Plague Inc.}{]} as a tool to teach the public
about outbreaks and disease transmission because of how it uses a
non-traditional route to raise public awareness on epidemiology, disease
transmission, and diseases/pandemic information'' \citep{kahn2013a}.
Potentially even more
exciting is to think about what the pattern methods offer if they were
adopted on a broad scale. Like \citet{escobar2018a} proposes, they could open
up the possibility to design multiple pluriverses, where we do not just
make plans focused on one choice but include diversity integrally in our
thinking. Reanalysis of futures methods in light of the
transdisciplinary pattern methodology has given us new insights into the
way futures studies work, and how they might be practiced
collaboratively, across domains and cultures.

\begin{figure}[h]
  \begin{center}
    \begin{mdframed}
      \begin{center}
\includegraphics[width=.5\columnwidth]{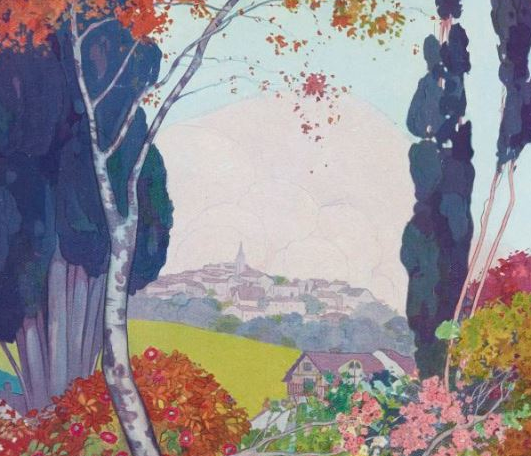}
      \end{center}
\begin{minipage}[t]{0.97\columnwidth}\raggedright
\subsection*{{\upshape{\texttt{EXAMPLES}}:}}

\textbf{Looking back: how }\emph{\textbf{might}}\textbf{ the patterns we
wrote have helped in 2020?}

\begin{itemize}
  \item In January, world leaders met online to simulate many global infection-spreading scenarios by playing Pandemic \citep{leacock2008a}, and Plague Inc \citep{creations2012a} to see how to contain or spread the virus.
  \item In February, people in nursing homes and prisons engaged in participatory scenario planning, which helped shift the collective outlook of society at large, leading to more informed discussions, and safer health behaviours.
  \item In March, national teams developed roadmaps for curricula teaching citizen science approaches of epidemiology, and students subsequently helped lead national recovery projects.
\end{itemize}

\textbf{Looking forward: How }\emph{\textbf{might}}\textbf{ things look
  in the future building on this work?}

\begin{itemize}
  \item Everyone learns “history” in school, but not “future”: this material and associated techniques could become part of a new curriculum for futures.
  \item Policy could develop via town-hall format discussions: rather than being based on Q\&A, politicians would lead citizen teams in developing different scenarios.
  \item E-sports could expand to explore plausible scenarios in games like SuperBugs \citep{nesta2016a} where players race against the clock to keep antibiotics effective.
  \end{itemize}
  
\end{minipage}
\end{mdframed}
\end{center}
\caption{Figure 5. Implications. Image: Umberto Brunelleschi, illustration pour
Fantasio d'Alfred de Musset, detail (Public Domain,
\href{https://commons.m.wikimedia.org/wiki/File:Umberto_Brunelleschi,_illustration_pour_Fantasio_d\%27Alfred_de_Musset.jpg}{\emph{via
Wikimedia Commons}})}
\end{figure}

\FloatBarrier

\section{Conclusions}

Our main results outlined several ways in which design patterns can be
used to model future-directed activities. It is possible to envision
scenarios that bring all of these methods together. For example,
consider a gaming situation in which teams are completing, or a similar
setting in which there are people acting with competing interests, with
different scenarios thrown into the mix
 (resource depletion, regulatory
changes, inflation, natural disasters, actions by other players, etc.)
to test their reactions. Roadmaps could be used both in the competition
and in research designs to make sense of the subsequent output and draw
transferable insights from the exercise.

As we have noted, a key feature of the methods we have described is to
promote heterogeneity of thought and behavior. Part of the purpose of
participatory methods would be to make sure we have the right experts.
For example, a nurse may have noticed things on the ward that
epidemiologists would never have experienced. A pandemic team with the
right balance of expertise and diverse community participants --- with
suitable methods for collaborating and learning together --- will get
better results than professional analysts, scientists, and university
researchers working in isolation.

Which brings us to you, dear reader. We assume you wanted to learn about
the subjects in the title: or perhaps this paper was assigned to you in
school and you are reading it for a grade? Either way, we are excited to
invite you to use and adapt these methods. For us, this work is likely
to serve as a springboard for our efforts. We hope that the paper will
be practical for you, too. Or, perhaps you think we are deluding
ourselves with ivory tower navel gazing? --- After all, even if the
broad outlines of this work are correct, a degree of skepticism is
warranted given that Alexander already shared relevant clues 20 years
ago. Or perhaps your criticisms run the other direction, and you found
the patterns too simplistic. We, of course, remain open to these and
other criticisms. Future work might integrate design patterns and
macrohistorical patterns for anticipation, or combine patterns and CLA
\citep{inayatullah1998a,inayatullah1998b} to connect diverse and divergent themes
across multiple layers of experience and observation. Alternative
methods described by \citet{Sools} focus on observing how consideration
of future outcomes affects present choices. Depending on what year it is
when you are reading this, perhaps these integrative suggestions have
already come to pass. If you think the points we raised in this paper
are meaningful and valuable, then a likely corollary is that there will
still be a lot of hard work to realise the full vision.

\section*{Acknowledgements}

We thank Claire van Rhyn for bringing the Anticipation conference to our
attention. We acknowledge the comments and participation in online
seminar discussions of: Roland Legrand, Lisa Snow MacDonald, Verena
Roberts, Charles Blass, Stephan Kreutzer, Giuliana Marques, and Cris
Gherhes.

\end{document}